# Do Water Molecules Always Stabilize Resonances? Microhydration Effects on Thymine Shape Resonances

Sujan Mandal[1], Jishnu Narayanan S J[1], Ankita Gogoi[1], Madhubani Mukherjee[2], Idan Haritan[3] and Achintya Kumar Dutta[1,*]

[1]Department of Chemistry, Indian Institute of Technology Bombay, Powai, Mumbai, 400076, India.

[2]Department of Chemistry, University of Southern California, Los Angeles, California, 90089-0482, United States

[3]Alexander Kofkin Faculty of Engineering, Bar-Ilan University, Ramat Gan, 5290002, Israel

**Abstract**

We investigate microhydration effects on the three low-lying π* shape resonances of thymine using the Resonance via Padé approach in combination with the DLPNO-EA-EOM-CCSD method. For isolated thymine, the calculated resonance positions are benchmarked against projected CAP-EA-EOM-CCSD calculations and compared with available theoretical and experimental data. Upon hydration, the 1π* and 2π* resonances undergo systematic stabilization accompanied by significant increase in their lifetimes, whereas the 3π* resonance exhibits a more complex behavior. In particular, the lifetime of the lowest resonance increases from 39 fs in isolated thymine to 110 fs in the thymine$(H_2O)_3$ cluster. Detailed analysis reveals that the observed resonance shifts arise from competing contributions involving hydrogen bonding, electrostatic interactions, microsolvation-induced geometric distortion, and finite-basis-set effects. Ghost-atom calculations demonstrate that diffuse basis functions associated with nearby water molecules contribute appreciably to the apparent stabilization, while explicit inclusion of water molecules leads to genuine physical stabilization of the resonance states. Furthermore, calculations on multiple conformers of the monohydrated cluster show that resonance positions and lifetimes depend strongly on the local hydrogen-bonding arrangement and microsolvation geometry. These findings demonstrate that resonance stabilization in microhydrated nucleobases is governed by a subtle interplay between geometry, basis-set effects, and intermolecular interactions.

## 1. Introduction

High-energy ionizing radiation (such as X-rays or γ-rays) is well known to damage biomolecules such as DNA and RNA. In recent decades, substantial experimental and theoretical efforts have been devoted to understanding the mechanisms of radiation-biomolecule interactions.[1–7] Radiation damage to DNA occurs through multiple pathways that are broadly classified as direct and indirect processes,[7–9] depending on the nature of the DNA-radiation interaction. In the direct pathway, high-energy photons interact directly with the genetic material, leading to ionization of DNA. Such direct damage can interfere with cellular division and proliferation and may ultimately result in cell death through necrosis or apoptosis. In contrast, indirect damage arises from reactive species, including reactive oxygen species and secondary electrons (SEs), that are primarily generated through water radiolysis.[10] In addition to these mechanisms, a quasi-direct pathway has also been proposed as an additional route to radiation-induced DNA damage.[7]

Although direct interactions between radiation and genetic material must be considered when discussing DNA damage caused by high-energy photons, a substantial fraction of the deposited radiation energy resides in the cellular environment surrounding DNA.[11] Ionization of the medium generates an array of reactive species, including SEs and free radicals, that can subsequently damage DNA.[12–21] Once generated, SEs lose energy through successive inelastic collisions with surrounding molecules, producing low-energy electrons (LEEs) that typically possess energies in the range of 0-20 eV. LEEs with energies below the ionization threshold of DNA (7.5-10 eV) can interact with biomolecules to form excited transient negative ions (TNIs) with lifetimes ranging from femtoseconds to picoseconds.[22–26] These TNIs, also referred to as anionic resonances, generally decay through three principal pathways:[27] (a) the additional electron can undergo autodetachment, resulting in the neutral molecule in the ground or an excited state, (b) the TNI can undergo geometric relaxation to form a stable radical anion, and (c) the dissociative electron attachment (DEA) pathway, which results in the fragmentation of the parent molecule to produce an anion and neutral radical species. DEA is of particular interest among the three pathways because of its involvement in LEE-induced radiation damage to genetic materials. DEA processes in DNA can give rise to single-strand breaks, double-strand breaks, cross-links, base release, and clustered damage sites.[7]

Numerous studies involving LEE irradiation of DNA subunits and oligonucleotides have highlighted the role of anionic resonance states in electron capture by DNA.[13–16,18,28–42] Incoming electrons can attach to DNA subunits (namely, the nucleobase, sugar, or phosphate groups) at specific energies to form TNIs.[30,43–46] Two main types of electronic resonances are observed in nucleobases:[8,47,48] shape resonances and Feshbach resonances. Shape resonances typically occur at lower energies and arise when the excess electron occupies a vacant molecular orbital. In contrast, Feshbach resonances involve simultaneous electron attachment and excitation of an electron within the molecules. Previous studies[8,49–53] have shown that both shape and Feshbach resonances can induce bond cleavage in DNA through DEA. For DEA to occur efficiently, the resonance lifetime must be sufficiently long for dissociation to compete with electron autodetachment. Therefore, investigating the characteristics of TNIs localized on nucleobases is essential for understanding DEA-induced DNA damage. Although both Feshbach and shape resonances may contribute to DNA damage, the present work focuses on nucleobase-centered shape resonances.

Electron attachment processes in biological systems occur primarily in the condensed phase, where water constitutes the dominant component of the surrounding environment. Therefore, studying the resonance states in an aqueous medium is essential to closely mimic the biological environment. Only a limited number of experimental studies have examined the condensed phase[54–56] effects on the resonance states of DNA nucleobases. Moreover, most theoretical studies of resonance states in DNA model systems have been performed in the gas phase,[47,57–61] with only a few investigations addressing aqueous environments.[62–65] Most of the studies have indicated that aqueous environments can stabilize shape resonances.[62–65] However, a recent investigation by Tripathi et al.[66] reported that bulk solvation leads to only modest stabilization of nucleobase-centered resonances on average, whereas significantly larger stabilization effects were observed in finite clusters. This apparent discrepancy highlights the need to disentangle the factors that contribute to the stabilization or destabilization of nucleobase-centered anionic resonances in aqueous environments. At the same time, explicit simulations of bulk water remain computationally demanding, highly sensitive to simulation details and sampling protocols, and often difficult to interpret in terms of a clear molecular picture. In this work, we use microsolvated thymine as a model system to isolate and quantify the competing effects of geometric distortion, intermolecular interactions, and finite basis set artifacts on nucleobase-centered resonance states.

## 2. Computational Details

Theoretical simulation of anionic resonances is more challenging than that of bound states because resonances are embedded in the continuum.[48] As a result, the corresponding wave functions are not square-integrable and cannot be represented as discrete eigenstates within the conventional Hermitian framework. Non-Hermitian quantum mechanics (NHQM) provides an alternative framework in which resonance states can be treated as discrete eigenstates within the time-independent Schrödinger equation formalism.[48] In this approach, the resonance energy can be described by the Siegert–Gamow equation:[67] $E = E_R - i\Gamma/2$ , where $E_R$ , the real part, describes the resonance energy (position) of the state and $\Gamma$ represents the resonance width, which is inversely related to the lifetime of the state. Various NHQM-based techniques have been developed for calculating resonances, including complex basis functions,[68–70] complex scaling,[71] and complex absorbing potential (CAP).[72–79] However, application of these methods generally requires significant modifications to conventional electronic structure codes. As an alternative, analytic continuation methods, which use analytic dilation of the energy from the real to the complex plane, can be used to study resonances without any modification of standard electronic structure codes. Among the available analytic continuation approaches, we employ the Resonance via Padé (RVP) method [80–83] to calculate resonance positions and widths. The RVP method relies on data obtained from a stabilization graph[84–86] (real or Hermitian plane), which is analytically continued to the complex plane (non-Hermitian regime) to extract resonance positions and widths. A detailed description of the RVP approach is available in the literature.[63,80,81,83,86]

The reliability of the calculated resonance positions and widths depends critically on the accuracy of the electronic structure method used to generate the stabilization graph. In this work, the electron-attachment equation-of-motion coupled-cluster (EA-EOM-CC) method[87–90] is employed to generate stabilization graphs for the systems studied. The EA-EOM-CC method is systematically improvable, yields size-intensive electron affinities, and is free from system-dependent parameterization. Additionally, it provides a balanced description of both dynamical and nondynamical electron correlation and enables access to multiple states within a single calculation.[72] In the present work, the EOM-CC framework is employed at the singles and doubles level (EOM-CCSD), which formally scales as O($N^6$) with system size. However, canonical EOM-CCSD becomes computationally prohibitive for larger systems due to its high computational cost and storage requirements.[91] The domain-based local pair natural orbital (DLPNO)[92] approximation to EOM-CCSD has proven highly effective for simulating ionization[93] and electron attachment processes,[91] as it provides a systematic balance between computational cost and accuracy. Previous studies have demonstrated that DLPNO-EA-EOM-CCSD provides reliable resonance energies for nucleobase systems.[63,65,94] In the present work, this method is employed to generate the stabilization graphs required for the RVP calculations. A detailed description of the theoretical framework and accuracy of the RVP-EA-EOM-DLPNO-CCSD method can be found in reference 63.

Initial geometries of microhydrated thymine complexes, (thymine$(H_2O)_n$, n= 1, 2, 3) were obtained via conformational sampling using the CREST package.[95] The lowest energy conformers were subsequently optimized at the RI-MP2/def2-TZVP level of theory using ORCA 5.0.3.[96] The geometry of the isolated neutral thymine was also optimized at the same level of theory. The optimized structures are shown in Figure 1, and the corresponding Cartesian coordinates are provided in the Supporting Information. Dunning's aug-cc-pVDZ basis set, augmented with additional diffuse 1s, 1p, and 1d functions, was employed for the EA-EOM-DLPNO-CCSD calculations. Additional functions were included solely on the heavy atoms, with exponents set to one-half of the exponent of the most diffuse functions of the corresponding angular momentum in the parent basis set. This modified basis set is denoted as aug-cc-pVDZ* throughout this work. To generate stabilization plots, the exponents of the two most diffuse s, p, and d functions on each atom were scaled by a factor $\alpha$ (except for hydrogen atoms, where only one diffuse function of each angular momentum was scaled), with $\alpha$ varied between 0.3 and 1.3. The stabilization curve for isolated thymine is shown in Figure 2, while those for all other systems are provided in the Supporting Information. The RVP calculations were performed using an open-source Python-based code, Automatic RVP.[97] For each resonance state, the reported energy and width correspond to the branch exhibiting the most statistically stable behavior. Additional calculations of resonance positions and widths for isolated thymine were performed using the projected-CAP-EA-EOM-CCSD[98,99] method with a box-type CAP potential and the aug-cc-pVDZ* basis set. The CAP onset parameters were set to 22.795, 17.148, and 6.068 Bohr along the X, Y, and Z directions, respectively. Projected CAP calculations were performed using Q-Chem[100] software package, while all other calculations were carried out using ORCA.[96]

## 3. Results and Discussion

### 3.1 Isolated thymine

Theoretical modeling of temporarily bound anions is challenging due to their strong sensitivity to both the basis set and the level of electronic structure theory. The reliability of the RVP-EA-EOM-DLPNO-CCSD approach (hereafter referred to as RVP) has been extensively examined in our previous studies.[63,65,94,101,102] Given the availability of both experimental and theoretical data for gas-phase thymine, we further benchmark our results against literature values. Isolated thymine exhibits three low-lying π* shape resonances. These are denoted as 1π*, 2π*, and 3π* based on the π* molecular orbitals involved in the initial electron attachment. The stabilization curve in Figure 2 highlights the stable regions corresponding to these three resonances. According to the RVP calculations (Table 1), the 1π* resonance appears at 0.67 eV with a lifetime of 39 fs, while the 2π* resonance appears at 2.42 eV with a lifetime of 11 fs. The 3π* resonance occurs at 5.38 eV and has a short lifetime of 6 fs. Additional projected box CAP-EA-EOM-CCSD[99] calculations were performed for isolated thymine to assess the accuracy of the RVP results. The resonance positions obtained using the RVP method are in reasonable agreement with the CAP-based results for all three states.

A similar trend is observed when our results are compared with those reported by Fennimore and Matsika.[47] In that work, the generalized Padé approximation (GPA) method, another stabilization-based approach, was employed to determine the resonance parameters of isolated thymine, with stabilization plots generated at the EA-EOM-CCSD/aug-cc-pVDZ* level of theory. The energies and widths of the 1π* and 2π* resonances obtained using RVP are in good agreement with the GPA results. In contrast, the 3π* resonance position obtained with RVP is higher by approximately 0.36 eV relative to the GPA value. This discrepancy likely reflects both methodological differences between the RVP and GPA approaches and the characteristics of the stabilization graph associated with the third resonance. In contrast to the first two resonances, the stabilization region associated with the third resonance is relatively steep. Because the RVP method utilizes the entire stable region as input, the resulting resonance position tends to lie between the corresponding avoided crossings. In contrast, the GPA method typically focuses on a single avoided crossing and therefore yields a resonance position closer to one of the limiting values, often the lower-energy value.

Additionally, $3\pi^*$ resonance may exhibit mixing with low-lying core-excited resonances[57], which are not adequately described at the present EOM-CCSD level of theory.

Experimental energies of thymine shape resonances were determined by Aflatooni et al.[103] using electron transmission spectroscopy (ETS). Representative theoretical results obtained using electron scattering methods,[59,104,105] CAP-SAC-CI,[75] and the stabilization approaches[47] are summarized in Table 1. A substantial variation exists in the resonance positions and widths reported using different theoretical methods. For the $1\pi^*$ resonance, reported positions span a wide range, from approximately 0.3 to 2.4 eV. Similar variability is observed for the higher resonances, with reported ranges of 1.9-5.5 eV for the $2\pi^*$ resonance and 5.02-7.9 eV for the $3\pi^*$ resonance. Compared with the ETS values,[103] theoretical resonance positions are generally overestimated. This trend is also observed for the present RVP results, likely due in part to the neglect of vibronic effects in the current calculations. However, the relative spacing between the resonance states obtained in this work is in reasonable agreement with experiment. The corresponding resonance widths exhibit even greater variability across different theoretical approaches.

### 3.2 Effect of aqueous media

The cellular environment can significantly influence the energy and lifetime of nucleobase-centered resonances of DNA. This, in turn, may affect the DEA-induced bond cleavage in the genetic material. To investigate these effects, we considered microhydrated thymine clusters, thymine$(H_2O)_n$ (n = 1-3). The qualitative character of the $1\pi^*$, $2\pi^*$, and $3\pi^*$ resonances remains unchanged upon hydration, as evidenced by the corresponding natural orbitals shown in Figure 3. This behavior contrasts with nucleobase-amino acid complexes, where additional resonances appear. This difference arises because amino acids can support their own low-lying anionic resonances; consequently, their inclusion introduces additional states that are either amino acid-centered or mixed in character. In particular, electron attachment to amino acids can generate low-lying $\pi^*$ type resonances localized primarily on the -COOH group.[102,106–108] In contrast, water molecules do not possess comparable low-lying resonances and therefore primarily perturb the nucleobase-centered resonances rather than generating new ones. The three resonance positions and corresponding widths for these thymine-water clusters are summarized in Table 2. Upon addition of a single water molecule, the $1\pi^*$ resonance position changes only marginally, decreasing by 0.01 eV. In contrast, the $2\pi^*$ resonance exhibits a more pronounced stabilization, decreasing by 0.11 eV. The widths of both resonances decrease, indicating longer lifetimes in the hydrated system. In contrast, the 3π* resonance exhibits a blue shift, with its position and width increasing by 0.08 and 0.04 eV, respectively.

These contrasting observations highlight the need to disentangle the factors governing resonance parameters under microsolvation. In addition to potentially stabilizing electrostatic interactions between water and the nucleobase, apparent shifts in resonance energies may also arise from basis-set effects and geometry changes induced by microsolvation. Diffuse basis functions on the water molecules can overlap with those on the solute, effectively increasing the flexibility of the basis set. To isolate these effects, we performed two additional RVP calculations (Table 3). In the first calculation, the water molecule was treated as ghost atoms, retaining its basis functions. In the second calculation, the water molecule was removed from thymine$(H_2O)_1$, leaving the nucleobase in the geometry adopted in the monohydrated complex. In the gas phase, thymine exhibits $C_s$ symmetry at its equilibrium geometry. Upon hydration, the thymine structure is distorted to $C_1$ symmetry in thymine$(H_2O)_1$. For simplicity, this distorted structure is referred to as thymine in $C_1$ geometry. Results in Table 3 show that all resonance positions exhibit a blue shift in the $C_1$ geometry relative to the equilibrium $C_s$ geometry. This indicates that geometric distortion and loss of symmetry in thymine have a destabilizing effect on all three resonance states. Notably, the 3π* state exhibits the largest destabilization (0.31 eV) due to geometric distortion. In the ghost-atom calculations for thymine$(H_2O)_1$, resonance positions decrease by 0.02, 0.02, and 0.12 eV for the 1π*, 2π*, and 3π* states, respectively, relative to the isolated thymine in the $C_1$ geometry. These apparent stabilizations can be attributed, at least in part, to finite basis set effects. Therefore, the stabilization observed for the 1π* and 2π* resonances in thymine$(H_2O)_1$ arises from a combination of electrostatic interactions and finite basis set effects.

As the number of water molecules increases from one to three, the energies of all three thymine shape resonances shift to lower values, and their lifetimes increase. However, the magnitude of the energy shift varies among the different resonance states. Upon addition of a second water molecule, the 1π* resonance shifts by -0.04 eV, whereas the 2π* resonance shifts by -0.19 eV relative to thymine$(H_2O)_1$. The 3π* resonance exhibits a red shift of 0.11 eV relative to thymine$(H_2O)_1$. Upon addition of a third water molecule, the 1π* and 2π* resonances further shift by -0.09 and -0.10 eV, respectively, relative to thymine$(H_2O)_2$. The 3π* resonance also exhibits a significant shift of -0.26 eV. In addition to these energy shifts, the resonance widths decrease for all states with an increasing number of water molecules. Compared with isolated thymine, the addition of three water molecules leads to a substantial increase in the lifetimes of the 1π* and 2π* resonances. The lifetime of the 1π* state increases from 39 to 110 fs, while that of the 2π* state increases from 11 to 44 fs. In contrast, the 3π* resonance exhibits only a modest increase in lifetime, from 6 to 8 fs. The significant increase in the 1π* resonance lifetime suggests that increasing microsolvation may eventually stabilize the anionic state.[65,109] However, direct extrapolation of the microsolvated results to a bound radical anion requires caution.[66]

To further examine the physical origin of resonance stabilization upon hydration, additional calculations were performed for thymine$(H_2O)_2$ and thymine$(H_2O)_3$ in which the water molecules were treated as ghost atoms (see Table 4). The results show that explicit inclusion of water molecules lowers all resonance energies relative to the corresponding ghost-atom calculations. For example, the 1π* resonance position in thymine$(H_2O)_3$ decreases from 0.61 eV in the ghost-atom calculation to 0.53 eV when water molecules are included explicitly. These results indicate that hydrogen-bonding and electrostatic interactions between water and thymine contribute to genuine physical stabilization of the resonance states in the microsolvated clusters.

### 3.3 Impact of the local geometry

The local geometry of the microsolvated cluster can also strongly influence electron attachment processes in nucleobases. Microhydrated nucleobases can adopt multiple conformations because of the presence of several hydrogen-bonding sites. The number of accessible conformers increases rapidly with the increasing number of solvent molecules. Three low-energy conformers of thymine$(H_2O)_1$ were considered, and their structures are shown in Figure S1 of the Supporting Information. The lowest-energy structure is denoted as Conformer-0, whereas Conformer-1 and Conformer-2 correspond to higher-energy structures arranged in order of increasing neutral-state energy. The corresponding results are summarized in Table 5. Natural-orbital analysis (Figure 4) reveals that the 1π* resonance state of Conformer-1 differs significantly from those of the other two conformers. In this conformer, one hydrogen atom of the water molecule is oriented toward a region of high electron density near the oxygen atom of thymine. This favorable interaction likely contributes to stronger stabilization of the 1π* resonance state, consistent with the values reported in Table 5. The calculated 1π* resonance position is 0.56 eV for Conformer-1, compared with 0.66 eV for the other two conformers. The 2π* resonance positions of Conformer-0 and Conformer-2 are nearly identical, appearing at 2.31 and 2.32 eV, respectively, with comparable resonance widths. In contrast, Conformer-1 exhibits a higher 2π* resonance energy of 2.42 eV. A similar trend is observed for the 3π* resonance, where Conformer-1 and Conformer-2 exhibit nearly identical resonance positions, while Conformer-0 appears at slightly higher energy. Figure 4 shows that the electron-density distribution near the water molecule is qualitatively similar for the 2π* and 3π* states across all conformers. The results suggest that the resonance positions and widths exhibit a strong dependence on the local environment in microhydrated clusters.

## 4. Conclusion

We have studied the effect of solvation on the shape resonance states of thymine, taking microhydrated thymine as a test case. The surrounding water molecules do not alter the qualitative nature of the resonance states but modify their positions and lifetimes. The addition of a single water molecule does not necessarily guarantee stabilization of all resonance states in the nucleobase. Using monohydrated thymine as an example, we demonstrate that several factors must be considered when analyzing these resonances. First, the geometric distortion induced by the addition of a water molecule leads to the destabilization of all three shape resonances in thymine. Second, finite basis set artifacts can lead to apparent stabilization of the resonance states. Consequently, the stabilization often observed for nucleobase shape resonances in the presence of a small number of water molecules may arise from basis set limitations, and meaningful conclusions about solvent effects cannot be drawn without correcting for basis set superposition errors. Finally, hydrogen-bonding interactions can lead to genuine stabilization of the resonance states. The interplay of these effects ultimately determines the behaviour of resonance states in an aqueous environment. Increasing the number of surrounding water molecules can lead to stabilization of the resonance states and an enhancement of their lifetimes. However, this stabilization does not follow a monotonic trend with solvation. In addition, the resonance positions and lifetimes depend strongly on the local microsolvation geometry, indicating substantial conformational sensitivity of nucleobase resonances in hydrated environments. Therefore, careful sampling of the solvent distribution around the nucleobase is essential to accurately describe the behaviour of nucleobase-centered resonance states in the bulk aqueous environment. Ongoing work is focused on addressing these aspects.

## Supporting Information

The Supporting Information contains the Cartesian coordinates and stabilization graphs for all systems considered in this study.

## Acknowledgement

The authors gratefully acknowledge financial support from IIT Bombay, including the IIT Bombay Seed Grant (Project No. R.D./0517-IRCCSH0-040), CRG (Project No. CRG/2022/005672), and MATRICS (Project No. MTR/2021/000420) projects of DST-SERB; CSIR-India (Project No. 01(3035)/21/EMR-II). The authors also acknowledge the IIT Bombay supercomputing facility and C-DAC resources (Param Smriti, Param Bramha, and Param Rudra) for providing computational time. S.M. gratefully acknowledges support from the Prime Minister's Research Fellowship (PMRF).

## Reference

(1) O'Neill, P. Radiation-Induced Damage in DNA. In *Studies in Physical and Theoretical Chemistry*; Jonah, C. D., Rao, B. S. M., Eds.; Radiation Chemistry; Elsevier, 2001; Vol. 87, pp 585–622. https://doi.org/10.1016/S0167-6881(01)80023-9.
(2) Cadet, J.; Bellon, S.; Douki, T.; Frelon, S.; Gasparutto, D.; Muller, E.; Pouget, J.-P.; Ravanat, J.-L.; Romieu, A.; Sauvaigo, S. Radiation-Induced DNA Damage: Formation, Measurement, and Biochemical Features. *J. Environ. Pathol. Toxicol. Oncol.* **2004**, *23* (1). https://doi.org/10.1615/JEnvPathToxOncol.v23.i1.30.
(3) Berthel, E.; Ferlazzo, M. L.; Devic, C.; Bourguignon, M.; Foray, N. What Does the History of Research on the Repair of DNA Double-Strand Breaks Tell Us?—A Comprehensive Review of Human Radiosensitivity. *Int. J. Mol. Sci.* **2019**, *20* (21), 5339. https://doi.org/10.3390/ijms20215339.
(4) Obodovskiy, I. *Radiation: Fundamentals, Applications, Risks, and Safety*; Elsevier, 2019.
(5) *DNA Damage, DNA Repair and Disease: Volume 1*; The Royal Society of Chemistry, 2020. https://doi.org/10.1039/9781839160769.
(6) *DNA Damage, DNA Repair and Disease: Volume 2*; The Royal Society of Chemistry, 2020. https://doi.org/10.1039/9781839162541.
(7) Narayanan S J, J.; Tripathi, D.; Verma, P.; Adhikary, A.; Dutta, A. K. Secondary Electron Attachment-Induced Radiation Damage to Genetic Materials. *ACS Omega* **2023**, *8* (12), 10669–10689. https://doi.org/10.1021/acsomega.2c06776.

(8) Alizadeh, E.; Orlando, T. M.; Sanche, L. Biomolecular Damage Induced by Ionizing Radiation: The Direct and Indirect Effects of Low-Energy Electrons on DNA. *Annu. Rev. Phys. Chem.* **2015**, *66*, 379–398. https://doi.org/10.1146/annurev-physchem-040513-103605.
(9) Kumar, A.; Becker, D.; Adhikary, A.; Sevilla, M. D. Reaction of Electrons with DNA: Radiation Damage to Radiosensitization. *Int. J. Mol. Sci.* **2019**, *20* (16), 3998. https://doi.org/10.3390/ijms20163998.
(10) Wang, H.; Mu, X.; He, H.; Zhang, X.-D. Cancer Radiosensitizers. *Trends Pharmacol. Sci.* **2018**, *39* (1), 24–48. https://doi.org/10.1016/j.tips.2017.11.003.
(11) Barcellos-Hoff, M. H.; Park, C.; Wright, E. G. Radiation and the Microenvironment – Tumorigenesis and Therapy. *Nat. Rev. Cancer* **2005**, *5* (11), 867–875. https://doi.org/10.1038/nrc1735.
(12) Boudaïffa, B.; Cloutier, P.; Hunting, D.; Huels, M. A.; Sanche, L. Resonant Formation of DNA Strand Breaks by Low-Energy (3 to 20 eV) Electrons. *Science* **2000**, *287* (5458), 1658–1660. https://doi.org/10.1126/science.287.5458.1658.
(13) Huels, M. A.; Boudaïffa, B.; Cloutier, P.; Hunting, D.; Sanche, L. Single, Double, and Multiple Double Strand Breaks Induced in DNA by 3−100 eV Electrons. *J. Am. Chem. Soc.* **2003**, *125* (15), 4467–4477. https://doi.org/10.1021/ja029527x.
(14) Abdoul-Carime, H.; Gohlke, S.; Fischbach, E.; Scheike, J.; Illenberger, E. Thymine Excision from DNA by Subexcitation Electrons. *Chem. Phys. Lett.* **2004**, *387* (4), 267–270. https://doi.org/10.1016/j.cplett.2004.02.022.
(15) Sanche, L. Low Energy Electron-Driven Damage in Biomolecules. *Eur. Phys. J. - At. Mol. Opt. Plasma Phys.* **2005**, *35* (2), 367–390. https://doi.org/10.1140/epjd/e2005-00206-6.
(16) Zheng, Y.; Cloutier, P.; Hunting, D. J.; Sanche, L.; Wagner, J. R. Chemical Basis of DNA Sugar−Phosphate Cleavage by Low-Energy Electrons. *J. Am. Chem. Soc.* **2005**, *127* (47), 16592–16598. https://doi.org/10.1021/ja054129q.
(17) Kopyra, J. Low Energy Electron Attachment to the Nucleotide Deoxycytidine Monophosphate: Direct Evidence for the Molecular Mechanisms of Electron-Induced DNA Strand Breaks. *Phys. Chem. Chem. Phys.* **2012**, *14* (23), 8287–8289. https://doi.org/10.1039/C2CP40847C.
(18) Khorsandgolchin, G.; Sanche, L.; Cloutier, P.; Wagner, J. R. Strand Breaks Induced by Very Low Energy Electrons: Product Analysis and Mechanistic Insight into the Reaction with TpT. *J. Am. Chem. Soc.* **2019**, *141* (26), 10315–10323. https://doi.org/10.1021/jacs.9b03295.
(19) Wang, C.-R.; Nguyen, J.; Lu, Q.-B. Bond Breaks of Nucleotides by Dissociative Electron Transfer of Nonequilibrium Prehydrated Electrons: A New Molecular Mechanism for Reductive DNA Damage. *J. Am. Chem. Soc.* **2009**, *131* (32), 11320–11322. https://doi.org/10.1021/ja902675g.
(20) Dong, Y.; Liao, H.; Gao, Y.; Cloutier, P.; Zheng, Y.; Sanche, L. Early Events in Radiobiology: Isolated and Cluster DNA Damage Induced by Initial Cations and Nonionizing Secondary Electrons. *J. Phys. Chem. Lett.* **2021**, *12* (1), 717–723. https://doi.org/10.1021/acs.jpclett.0c03341.
(21) Narayanan S J, J.; Tripathi, D.; Dutta, A. K. Doorway Mechanism for Electron Attachment Induced DNA Strand Breaks. *J. Phys. Chem. Lett.* **2021**, *12* (42), 10380–10387. https://doi.org/10.1021/acs.jpclett.1c02735.
(22) Herbert, J. M.; Coons, M. P. The Hydrated Electron. *Annu. Rev. Phys. Chem.* **2017**, *68*, 447–472. https://doi.org/10.1146/annurev-physchem-052516-050816.

(23) Alizadeh, E.; Sanche, L. Precursors of Solvated Electrons in Radiobiological Physics and Chemistry. *Chem. Rev.* **2012**, *112* (11), 5578–5602. https://doi.org/10.1021/cr300063r.
(24) Walker, D. C. The Hydrated Electron. *Q. Rev. Chem. Soc.* **1967**, *21* (1), 79–108. https://doi.org/10.1039/QR9672100079.
(25) Kumar, A.; Sevilla, M. D. Low-Energy Electron (LEE)-Induced DNA Damage: Theoretical Approaches to Modeling Experiment. In *Handbook of Computational Chemistry*; Leszczynski, J., Ed.; Springer Netherlands: Dordrecht, 2016; pp 1–63. https://doi.org/10.1007/978-94-007-6169-8_34-2.
(26) Ma, J.; Kumar, A.; Muroya, Y.; Yamashita, S.; Sakurai, T.; Denisov, S. A.; Sevilla, M. D.; Adhikary, A.; Seki, S.; Mostafavi, M. Observation of Dissociative Quasi-Free Electron Attachment to Nucleoside via Excited Anion Radical in Solution. *Nat. Commun.* **2019**, *10* (1), 102. https://doi.org/10.1038/s41467-018-08005-z.
(27) Arumainayagam, C. R.; Lee, H.-L.; Nelson, R. B.; Haines, D. R.; Gunawardane, R. P. Low-Energy Electron-Induced Reactions in Condensed Matter. *Surf. Sci. Rep.* **2010**, *65*, 1–44. https://doi.org/10.1016/j.surfrep.2009.09.001.
(28) Kumari, B.; Huwaidi, A.; Robert, G.; Cloutier, P.; Bass, A. D.; Sanche, L.; Wagner, J. R. Shape Resonances in DNA: Nucleobase Release, Reduction, and Dideoxynucleoside Products Induced by 1.3 to 2.3 eV Electrons. *J. Phys. Chem. B* **2022**, *126* (28), 5175–5184. https://doi.org/10.1021/acs.jpcb.2c01851.
(29) Adjei, D.; de Cabrera, M.; Reyes, Y.; Barbolovici, A.; Alahmadi, M.; Ward, S.; Menet, M.-C.; Mejanelle, P.; Kumar, A.; Sevilla, M. D.; Wnuk, S. F.; Mostafavi, M.; Adhikary, A. The Site of Azido Substitution in a Pyrimidine Nucleobase Dictates the Type of Nitrogen-Centered Radical Formed after Dissociative Electron Attachment. *J. Phys. Chem. B* **2025**, *129* (32), 8115–8126. https://doi.org/10.1021/acs.jpcb.5c02751.
(30) Barrios, R.; Skurski, P.; Simons, J. Mechanism for Damage to DNA by Low-Energy Electrons. *J. Phys. Chem. B* **2002**, *106* (33), 7991–7994. https://doi.org/10.1021/jp013861i.
(31) Pan, X.; Cloutier, P.; Hunting, D.; Sanche, L. Dissociative Electron Attachment to DNA. *Phys. Rev. Lett.* **2003**, *90* (20), 208102. https://doi.org/10.1103/PhysRevLett.90.208102.
(32) Simons, J. How Do Low-Energy (0.1−2 eV) Electrons Cause DNA-Strand Breaks? *Acc. Chem. Res.* **2006**, *39* (10), 772–779. https://doi.org/10.1021/ar0680769.
(33) Gu, J.; Xie, Y.; Schaefer, H. F. Near 0 eV Electrons Attach to Nucleotides. *J. Am. Chem. Soc.* **2006**, *128* (4), 1250–1252. https://doi.org/10.1021/ja055615g.
(34) Smyth, M.; Kohanoff, J. Excess Electron Interactions with Solvated DNA Nucleotides: Strand Breaks Possible at Room Temperature. *J. Am. Chem. Soc.* **2012**, *134* (22), 9122–9125. https://doi.org/10.1021/ja303776r.
(35) Kumar, A.; Sevilla, M. D. Role of Excited States in Low-Energy Electron (LEE) Induced Strand Breaks in DNA Model Systems: Influence of Aqueous Environment. *ChemPhysChem* **2009**, *10* (9–10), 1426–1430. https://doi.org/10.1002/cphc.200900025.
(36) Bhaskaran, R.; Sarma, M. The Role of the Shape Resonance State in Low Energy Electron Induced Single Strand Break in 2′-Deoxycytidine-5′-Monophosphate. *Phys. Chem. Chem. Phys.* **2015**, *17* (23), 15250–15257. https://doi.org/10.1039/C5CP00126A.
(37) Bhaskaran, R.; Sarma, M. Low Energy Electron Induced Cytosine Base Release in 2′-Deoxycytidine-3′-Monophosphate via Glycosidic Bond Cleavage: A Time-Dependent Wavepacket Study. *J. Chem. Phys.* **2014**, *141* (10), 104309. https://doi.org/10.1063/1.4894744.

(38) McAllister, M.; Smyth, M.; Gu, B.; Tribello, G. A.; Kohanoff, J. Understanding the Interaction between Low-Energy Electrons and DNA Nucleotides in Aqueous Solution. *J. Phys. Chem. Lett.* **2015**, *6* (15), 3091–3097. https://doi.org/10.1021/acs.jpclett.5b01011.
(39) McAllister, M.; Kazemigazestane, N.; Henry, L. T.; Gu, B.; Fabrikant, I.; Tribello, G. A.; Kohanoff, J. Solvation Effects on Dissociative Electron Attachment to Thymine. *J. Phys. Chem. B* **2019**, *123* (7), 1537–1544. https://doi.org/10.1021/acs.jpcb.8b11621.
(40) Kumar, S.; Sarmah, M. P.; Dash, P. S.; Sarma, M. Study of Electron Capture by 5-Halogenated Cytosine in Gas and Condensed Phases. *J. Phys. Chem. A* **2025**, *129* (32), 7429–7439. https://doi.org/10.1021/acs.jpca.5c04881.
(41) Davis, D.; Sajeev, Y. A Hitherto Unknown Stability of DNA Basepairs. *Chem. Commun.* **2020**, *56* (93), 14625–14628. https://doi.org/10.1039/D0CC06641A.
(42) Anstöter, C. S.; DelloStritto, M.; Klein, M. L.; Matsika, S. Modeling the Ultrafast Electron Attachment Dynamics of Solvated Uracil. *J. Phys. Chem. A* **2021**, *125* (32), 6995–7003. https://doi.org/10.1021/acs.jpca.1c05288.
(43) Li, X.; Sevilla, M. D.; Sanche, L. Density Functional Theory Studies of Electron Interaction with DNA: Can Zero eV Electrons Induce Strand Breaks? *J. Am. Chem. Soc.* **2003**, *125* (45), 13668–13669. https://doi.org/10.1021/ja036509m.
(44) Ptasińska, S.; Denifl, S.; Scheier, P.; Märk, T. D. Inelastic Electron Interaction (Attachment/Ionization) with Deoxyribose. *J. Chem. Phys.* **2004**, *120* (18), 8505–8511. https://doi.org/10.1063/1.1690231.
(45) Sommerfeld, T. Doorway Mechanism for Dissociative Electron Attachment to Fructose. *J. Chem. Phys.* **2007**, *126* (12), 124301. https://doi.org/10.1063/1.2710275.
(46) Baccarelli, I.; Gianturco, F. A.; Grandi, A.; Sanna, N. Metastable Anion Fragmentations after Resonant Attachment: Deoxyribosic Structures from Quantum Electron Dynamics. *Int. J. Quantum Chem.* **2008**, *108* (11), 1878–1887. https://doi.org/10.1002/qua.21681.
(47) Fennimore, M. A.; Matsika, S. Electronic Resonances of Nucleobases Using Stabilization Methods. *J. Phys. Chem. A* **2018**, *122* (16), 4048–4057. https://doi.org/10.1021/acs.jpca.8b01523.
(48) Moiseyev, N. *Non-Hermitian Quantum Mechanics*; Cambridge University Press, 2011.
(49) Martin, F. DNA Strand Breaks Induced by 0–4 eV Electrons: The Role of Shape Resonances. *Phys. Rev. Lett.* **2004**, *93* (6), 068101. https://doi.org/10.1103/PhysRevLett.93.068101.
(50) Berdys, J.; Anusiewicz, I.; Skurski, P.; Simons, J. Damage to Model DNA Fragments from Very Low-Energy (<1 eV) Electrons. *J. Am. Chem. Soc.* **2004**, *126* (20), 6441–6447. https://doi.org/10.1021/ja049876m.
(51) Panajotovic, R.; Martin, F.; Cloutier, P.; Hunting, D.; Sanche, L. Effective Cross Sections for Production of Single-Strand Breaks in Plasmid DNA by 0.1 to 4.7 eV Electrons. *Radiat. Res.* **2006**, *165* (4), 452–459. https://doi.org/10.1667/RR3521.1.
(52) Zheng, Y.; Cloutier, P.; Hunting, D. J.; Wagner, J. R.; Sanche, L. Phosphodiester and N-Glycosidic Bond Cleavage in DNA Induced by 4–15 eV Electrons. *J. Chem. Phys.* **2006**, *124* (6), 064710. https://doi.org/10.1063/1.2166364.
(53) Luo, X.; Zheng, Y.; Sanche, L. DNA Strand Breaks and Crosslinks Induced by Transient Anions in the Range 2-20 eV. *J. Chem. Phys.* **2014**, *140* (15), 155101. https://doi.org/10.1063/1.4870519.

(54) Cooper, G. A.; Clarke, C. J.; Verlet, J. R. R. Low-Energy Shape Resonances of a Nucleobase in Water. *J. Am. Chem. Soc.* **2023**, *145* (2), 1319–1326. https://doi.org/10.1021/jacs.2c11440.
(55) Clarke, C. J.; Burrow, E. M.; Verlet, J. R. R. The Influence of Water Molecules on the Π* Shape Resonances of the Thymine Anion. *J. Phys. Chem. A* **2025**, *129* (26), 5771–5778. https://doi.org/10.1021/acs.jpca.5c01948.
(56) Kočišek, J.; Pysanenko, A.; Fárník, M.; Fedor, J. Microhydration Prevents Fragmentation of Uracil and Thymine by Low-Energy Electrons. *J. Phys. Chem. Lett.* **2016**, *7* (17), 3401–3405. https://doi.org/10.1021/acs.jpclett.6b01601.
(57) Fennimore, M. A.; Matsika, S. Core-Excited and Shape Resonances of Uracil. *Phys. Chem. Chem. Phys.* **2016**, *18* (44), 30536–30545. https://doi.org/10.1039/C6CP05342D.
(58) Winstead, C.; McKoy, V. Low-Energy Electron Collisions with Gas-Phase Uracil. *J. Chem. Phys.* **2006**, *125* (17), 174304. https://doi.org/10.1063/1.2353147.
(59) Dora, A.; Bryjko, L.; Mourik, T. van; Tennyson, J. R-Matrix Study of Elastic and Inelastic Electron Collisions with Cytosine and Thymine. *J. Phys. B: At. Mol. Opt. Phys.* **2012**, *45* (17), 175203. https://doi.org/10.1088/0953-4075/45/17/175203.
(60) Dora, A.; Tennyson, J.; Bryjko, L.; Mourik, T. van. R-Matrix Calculation of Low-Energy Electron Collisions with Uracil. *J. Chem. Phys.* **2009**, *130* (16), 164307. https://doi.org/10.1063/1.3119667.
(61) Bouskila, G.; Landau, A.; Haritan, I.; Moiseyev, N.; Bhattacharya, D. Complex Energies and Transition Dipoles for Shape-Type Resonances of Uracil Anion from Stabilization Curves via Padé. *J. Chem. Phys.* **2022**, *156* (19), 194101. https://doi.org/10.1063/5.0086887.
(62) Sieradzka, A.; Gorfinkiel, J. D. Theoretical Study of Resonance Formation in Microhydrated Molecules. II. Thymine-(H2O)n, n = 1,2,3,5. *J. Chem. Phys.* **2017**, *147* (3), 034303. https://doi.org/10.1063/1.4993946.
(63) Verma, P.; Mukherjee, M.; Bhattacharya, D.; Haritan, I.; Dutta, A. K. Shape Resonance Induced Electron Attachment to Cytosine: The Effect of Aqueous Media. *J. Chem. Phys.* **2023**, *159* (21), 214303. https://doi.org/10.1063/5.0157576.
(64) Cornetta, L. M.; Coutinho, K.; Varella, M. T. do N. Solvent Effects on the Π* Shape Resonances of Uracil. *J. Chem. Phys.* **2020**, *152* (8), 084301. https://doi.org/10.1063/1.5139459.
(65) Narayanan S J, J.; Tripathi, D.; Haritan, I.; Dutta, A. K. The Effect of Aqueous Medium on Nucleobase Shape Resonances: Insights from Microsolvation. *J. Phys. Chem. A* **2025**, *129* (48), 11179–11188. https://doi.org/10.1021/acs.jpca.5c06868.
(66) Tripathi, D.; Pyla, M.; Dutta, A. K.; Matsika, S. Impact of Solvation on the Electronic Resonances in Uracil. *Phys. Chem. Chem. Phys.* **2025**, *27* (7), 3588–3601. https://doi.org/10.1039/D4CP04333B.
(67) Siegert, A. J. F. On the Derivation of the Dispersion Formula for Nuclear Reactions. *Phys. Rev.* **1939**, *56* (8), 750–752. https://doi.org/10.1103/PhysRev.56.750.
(68) Rescigno, T. N.; McCurdy, C. W.; Orel, A. E. Extensions of the Complex-Coordinate Method to the Study of Resonances in Many-Electron Systems. *Phys. Rev. A* **1978**, *17* (6), 1931–1938. https://doi.org/10.1103/PhysRevA.17.1931.
(69) Hernández Vera, M.; Jagau, T.-C. Resolution-of-the-Identity Second-Order Møller–Plesset Perturbation Theory with Complex Basis Functions: Benchmark Calculations and Applications to Strong-Field Ionization of Polyacenes. *J. Chem. Phys.* **2020**, *152* (17), 174103. https://doi.org/10.1063/5.0004843.

(70) White, A. F.; Epifanovsky, E.; McCurdy, C. W.; Head-Gordon, M. Second Order Møller-Plesset and Coupled Cluster Singles and Doubles Methods with Complex Basis Functions for Resonances in Electron-Molecule Scattering. *J. Chem. Phys.* **2017**, *146* (23), 234107. https://doi.org/10.1063/1.4986950.
(71) Moiseyev, N.; Corcoran, C. Autoionizing States of $H_2$ and $H_2^-$ Using the Complex-Scaling Method. *Phys. Rev. A* **1979**, *20* (3), 814–817. https://doi.org/10.1103/PhysRevA.20.814.
(72) Jagau, T.-C.; Bravaya, K. B.; Krylov, A. I. Extending Quantum Chemistry of Bound States to Electronic Resonances. *Annu. Rev. Phys. Chem.* **2017**, *68*, 525–553. https://doi.org/10.1146/annurev-physchem-052516-050622.
(73) Ghosh, A.; Vaval, N.; Pal, S. Equation-of-Motion Coupled-Cluster Method for the Study of Shape Resonance. *J. Chem. Phys.* **2012**, *136* (23), 234110. https://doi.org/10.1063/1.4729464.
(74) Santra, R.; Cederbaum, L. S. Non-Hermitian Electronic Theory and Applications to Clusters. *Phys. Rep.* **2002**, *368* (1), 1–117. https://doi.org/10.1016/S0370-1573(02)00143-6.
(75) Kanazawa, Y.; Ehara, M.; Sommerfeld, T. Low-Lying Π* Resonances of Standard and Rare DNA and RNA Bases Studied by the Projected CAP/SAC–CI Method. *J. Phys. Chem. A* **2016**, *120* (9), 1545–1553. https://doi.org/10.1021/acs.jpca.5b12190.
(76) Sajeev, Y.; Ghosh, A.; Vaval, N.; Pal, S. Coupled Cluster Methods for Autoionisation Resonances. *Int. Rev. Phys. Chem.* **2014**, *33* (3), 397–425. https://doi.org/10.1080/0144235X.2014.935585.
(77) Jagau, T.-C.; Zuev, D.; Bravaya, K. B.; Epifanovsky, E.; Krylov, A. I. A Fresh Look at Resonances and Complex Absorbing Potentials: Density Matrix-Based Approach. *J. Phys. Chem. Lett.* **2014**, *5* (2), 310–315. https://doi.org/10.1021/jz402482a.
(78) Sajeev, Y.; Sindelka, M.; Moiseyev, N. Reflection-Free Complex Absorbing Potential for Electronic Structure Calculations: Feshbach Type Autoionization Resonance of Helium. *Chem. Phys.* **2006**, *329* (1), 307–312. https://doi.org/10.1016/j.chemphys.2006.08.008.
(79) Sajeev, Y.; Moiseyev, N. Reflection-Free Complex Absorbing Potential for Electronic Structure Calculations: Feshbach-Type Autoionization Resonances of Molecules. *J. Chem. Phys.* **2007**, *127* (3), 034105. https://doi.org/10.1063/1.2753485.
(80) Landau, A.; Haritan, I. The Clusterization Technique: A Systematic Search for the Resonance Energies Obtained via Padé. *J. Phys. Chem. A* **2019**, *123* (24), 5091–5105. https://doi.org/10.1021/acs.jpca.8b12573.
(81) Landau, A.; Haritan, I.; Kaprálová-Žd'ánská, P. R.; Moiseyev, N. Atomic and Molecular Complex Resonances from Real Eigenvalues Using Standard (Hermitian) Electronic Structure Calculations. *J. Phys. Chem. A* **2016**, *120* (19), 3098–3108. https://doi.org/10.1021/acs.jpca.5b10685.
(82) Haritan, I.; Moiseyev, N. On the Calculation of Resonances by Analytic Continuation of Eigenvalues from the Stabilization Graph. *J. Chem. Phys.* **2017**, *147* (1), 014101. https://doi.org/10.1063/1.4989867.
(83) Landau, A.; Haritan, I.; Moiseyev, N. The RVP Method From Real Ab-Initio Calculations to Complex Energies and Transition Dipoles. *Front. Phys.* **2022**, *10*. https://doi.org/10.3389/fphy.2022.854039.
(84) Holøien, E.; Midtdal, J. New Investigation of the 1SeAutoionizing States of He and H−. *J. Chem. Phys.* **1966**, *45* (6), 2209–2216. https://doi.org/10.1063/1.1727912.
(85) Hazi, A. U.; Taylor, H. S. Stabilization Method of Calculating Resonance Energies: Model Problem. *Phys. Rev. A* **1970**, *1* (4), 1109–1120. https://doi.org/10.1103/PhysRevA.1.1109.

(86) Landau, A. Shaping and Controlling Stabilisation Graphs for Calculating Stable Complex Resonance Energies. *Mol. Phys.* **2019**, *117* (15–16), 2029–2042. https://doi.org/10.1080/00268976.2019.1575993.
(87) Stanton, J. F.; Bartlett, R. J. The Equation of Motion Coupled-cluster Method. A Systematic Biorthogonal Approach to Molecular Excitation Energies, Transition Probabilities, and Excited State Properties. *J. Chem. Phys.* **1993**, *98* (9), 7029–7039. https://doi.org/10.1063/1.464746.
(88) Krylov, A. I. Equation-of-Motion Coupled-Cluster Methods for Open-Shell and Electronically Excited Species: The Hitchhiker's Guide to Fock Space. *Annu. Rev. Phys. Chem.* **2008**, *59*, 433–462. https://doi.org/10.1146/annurev.physchem.59.032607.093602.
(89) Kowalski, K.; Piecuch, P. The Active-Space Equation-of-Motion Coupled-Cluster Methods for Excited Electronic States: Full EOMCCSDt. *J. Chem. Phys.* **2001**, *115* (2), 643–651. https://doi.org/10.1063/1.1378323.
(90) Nooijen, M.; Bartlett, R. J. Equation of Motion Coupled Cluster Method for Electron Attachment. *J. Chem. Phys.* **1995**, *102* (9), 3629–3647. https://doi.org/10.1063/1.468592.
(91) Dutta, A. K.; Saitow, M.; Demoulin, B.; Neese, F.; Izsák, R. A Domain-Based Local Pair Natural Orbital Implementation of the Equation of Motion Coupled Cluster Method for Electron Attached States. *J. Chem. Phys.* **2019**, *150* (16), 164123. https://doi.org/10.1063/1.5089637.
(92) Riplinger, C.; Neese, F. An Efficient and near Linear Scaling Pair Natural Orbital Based Local Coupled Cluster Method. *J. Chem. Phys.* **2013**, *138* (3), 034106. https://doi.org/10.1063/1.4773581.
(93) Dutta, A. K.; Saitow, M.; Riplinger, C.; Neese, F.; Izsák, R. A Near-Linear Scaling Equation of Motion Coupled Cluster Method for Ionized States. *J. Chem. Phys.* **2018**, *148* (24), 244101. https://doi.org/10.1063/1.5029470.
(94) Narayanan S J, J.; Tripathi, D.; Haritan, I.; Adhikary, A.; Pandey, B.; Dutta, A. K. The Effect of Base-Pairing on the Shape Resonances of Nucleobases. *J. Chem. Phys.* **2026**, *164* (18), 184301. https://doi.org/10.1063/5.0333805.
(95) Pracht, P.; Bohle, F.; Grimme, S. Automated Exploration of the Low-Energy Chemical Space with Fast Quantum Chemical Methods. *Phys. Chem. Chem. Phys.* **2020**, *22* (14), 7169–7192. https://doi.org/10.1039/C9CP06869D.
(96) Neese, F. Software Update: The ORCA Program System—Version 5.0. *WIREs Comput. Mol. Sci.* **2022**, *12* (5), e1606. https://doi.org/10.1002/wcms.1606.
(97) Automatic-RVP: GitHub repository. https://github.com/haritan/RVP (accessed 2026-05-10).
(98) Zuev, D.; Jagau, T.-C.; Bravaya, K. B.; Epifanovsky, E.; Shao, Y.; Sundstrom, E.; Head-Gordon, M.; Krylov, A. I. Complex Absorbing Potentials within EOM-CC Family of Methods: Theory, Implementation, and Benchmarks. *J. Chem. Phys.* **2014**, *141* (2), 024102. https://doi.org/10.1063/1.4885056.
(99) Gayvert, J. R.; Bravaya, K. B. Projected CAP-EOM-CCSD Method for Electronic Resonances. *J. Chem. Phys.* **2022**, *156* (9), 094108. https://doi.org/10.1063/5.0082739.
(100) Epifanovsky, E.; Gilbert, A. T. B.; Feng, X.; Lee, J.; Mao, Y.; Mardirossian, N.; Pokhilko, P.; White, A. F.; Coons, M. P.; Dempwolff, A. L.; Gan, Z.; Hait, D.; Horn, P. R.; Jacobson, L. D.; Kaliman, I.; Kussmann, J.; Lange, A. W.; Lao, K. U.; Levine, D. S.; Liu, J.; McKenzie, S. C.; Morrison, A. F.; Nanda, K. D.; Plasser, F.; Rehn, D. R.; Vidal, M. L.; You, Z.-Q.; Zhu, Y.; Alam, B.; Albrecht, B. J.; Aldossary, A.; Alguire, E.; Andersen, J. H.; Athavale, V.; Barton,

D.; Begam, K.; Behn, A.; Bellonzi, N.; Bernard, Y. A.; Berquist, E. J.; Burton, H. G. A.; Carreras, A.; Carter-Fenk, K.; Chakraborty, R.; Chien, A. D.; Closser, K. D.; Cofer-Shabica, V.; Dasgupta, S.; de Wergifosse, M.; Deng, J.; Diedenhofen, M.; Do, H.; Ehlert, S.; Fang, P.-T.; Fatehi, S.; Feng, Q.; Friedhoff, T.; Gayvert, J.; Ge, Q.; Gidofalvi, G.; Goldey, M.; Gomes, J.; González-Espinoza, C. E.; Gulania, S.; Gunina, A. O.; Hanson-Heine, M. W. D.; Harbach, P. H. P.; Hauser, A.; Herbst, M. F.; Hernández Vera, M.; Hodecker, M.; Holden, Z. C.; Houck, S.; Huang, X.; Hui, K.; Huynh, B. C.; Ivanov, M.; Jász, Á.; Ji, H.; Jiang, H.; Kaduk, B.; Kähler, S.; Khistyaev, K.; Kim, J.; Kis, G.; Klunzinger, P.; Koczor-Benda, Z.; Koh, J. H.; Kosenkov, D.; Koulias, L.; Kowalczyk, T.; Krauter, C. M.; Kue, K.; Kunitsa, A.; Kus, T.; Ladjánszki, I.; Landau, A.; Lawler, K. V.; Lefrancois, D.; Lehtola, S.; Li, R. R.; Li, Y.-P.; Liang, J.; Liebenthal, M.; Lin, H.-H.; Lin, Y.-S.; Liu, F.; Liu, K.-Y.; Loipersberger, M.; Luenser, A.; Manjanath, A.; Manohar, P.; Mansoor, E.; Manzer, S. F.; Mao, S.-P.; Marenich, A. V.; Markovich, T.; Mason, S.; Maurer, S. A.; McLaughlin, P. F.; Menger, M. F. S. J.; Mewes, J.-M.; Mewes, S. A.; Morgante, P.; Mullinax, J. W.; Oosterbaan, K. J.; Paran, G.; Paul, A. C.; Paul, S. K.; Pavošević, F.; Pei, Z.; Prager, S.; Proynov, E. I.; Rák, Á.; Ramos-Cordoba, E.; Rana, B.; Rask, A. E.; Rettig, A.; Richard, R. M.; Rob, F.; Rossomme, E.; Scheele, T.; Scheurer, M.; Schneider, M.; Sergueev, N.; Sharada, S. M.; Skomorowski, W.; Small, D. W.; Stein, C. J.; Su, Y.-C.; Sundstrom, E. J.; Tao, Z.; Thirman, J.; Tornai, G. J.; Tsuchimochi, T.; Tubman, N. M.; Veccham, S. P.; Vydrov, O.; Wenzel, J.; Witte, J.; Yamada, A.; Yao, K.; Yeganeh, S.; Yost, S. R.; Zech, A.; Zhang, I. Y.; Zhang, X.; Zhang, Y.; Zuev, D.; Aspuru-Guzik, A.; Bell, A. T.; Besley, N. A.; Bravaya, K. B.; Brooks, B. R.; Casanova, D.; Chai, J.-D.; Coriani, S.; Cramer, C. J.; Cserey, G.; DePrince, A. E., III; DiStasio, R. A., Jr.; Dreuw, A.; Dunietz, B. D.; Furlani, T. R.; Goddard, W. A., III; Hammes-Schiffer, S.; Head-Gordon, T.; Hehre, W. J.; Hsu, C.-P.; Jagau, T.-C.; Jung, Y.; Klamt, A.; Kong, J.; Lambrecht, D. S.; Liang, W.; Mayhall, N. J.; McCurdy, C. W.; Neaton, J. B.; Ochsenfeld, C.; Parkhill, J. A.; Peverati, R.; Rassolov, V. A.; Shao, Y.; Slipchenko, L. V.; Stauch, T.; Steele, R. P.; Subotnik, J. E.; Thom, A. J. W.; Tkatchenko, A.; Truhlar, D. G.; Van Voorhis, T.; Wesolowski, T. A.; Whaley, K. B.; Woodcock, H. L., III; Zimmerman, P. M.; Faraji, S.; Gill, P. M. W.; Head-Gordon, M.; Herbert, J. M.; Krylov, A. I. Software for the Frontiers of Quantum Chemistry: An Overview of Developments in the Q-Chem 5 Package. *J. Chem. Phys.* **2021**, *155* (8), 084801. https://doi.org/10.1063/5.0055522.

(101) Arora, S.; Narayanan, S. J. J.; Dutta, A. K. How Good Is the Time-Dependent DFT Method for Simulating Anionic Shape Resonances of DNA Nucleobases? *J. Chem. Sci.* **2025**, *137* (4), 119. https://doi.org/10.1007/s12039-025-02451-1.

(102) Arora, S.; Narayanan S J, J.; Haritan, I.; Adhikary, A.; Dutta, A. K. Effect of Protein Environment on the Shape Resonances of RNA Pyrimidine Nucleobases: Insights from a Model System. *J. Chem. Phys.* **2025**, *163* (13), 134103. https://doi.org/10.1063/5.0288514.

(103) Aflatooni, K.; Gallup, G. A.; Burrow, P. D. Electron Attachment Energies of the DNA Bases. *J. Phys. Chem. A* **1998**, *102* (31), 6205–6207. https://doi.org/10.1021/jp980865n.

(104) Tonzani, S.; Greene, C. H. Low-Energy Electron Scattering from DNA and RNA Bases: Shape Resonances and Radiation Damage. *J. Chem. Phys.* **2006**, *124* (5), 054312. https://doi.org/10.1063/1.2148965.

(105) Winstead, C.; McKoy, V.; d'Almeida Sanchez, S. Interaction of Low-Energy Electrons with the Pyrimidine Bases and Nucleosides of DNA. *J. Chem. Phys.* **2007**, *127* (8), 085105. https://doi.org/10.1063/1.2757617.

(106) Kumar, S.; Singh, H. K.; Bhattacharyya, H. P.; Sarma, M. Low Energy Electron Interaction with Citric Acid: A Local Complex Potential Based Time-Dependent Wavepacket Study. *J. Chem. Sci.* **2023**, *135* (3), 88. https://doi.org/10.1007/s12039-023-02200-2.
(107) Sarmah, M. P.; Medhi, B.; Sarma, M. Impact of the Electron Attachment to the Alanyl Glycine and Glycyl Alanine Conformers. *J. Chem. Phys.* **2026**, *164* (5), 054307. https://doi.org/10.1063/5.0297297.
(108) Aflatooni, K.; Hitt, B.; Gallup, G. A.; Burrow, P. D. Temporary Anion States of Selected Amino Acids. *J. Chem. Phys.* **2001**, *115* (14), 6489–6494. https://doi.org/10.1063/1.1404147.
(109) Mukherjee, M.; Tripathi, D.; Dutta, A. K. Water Mediated Electron Attachment to Nucleobases: Surface-Bound vs Bulk Solvated Electrons. *J. Chem. Phys.* **2020**, *153* (4), 044305. https://doi.org/10.1063/5.0010509.

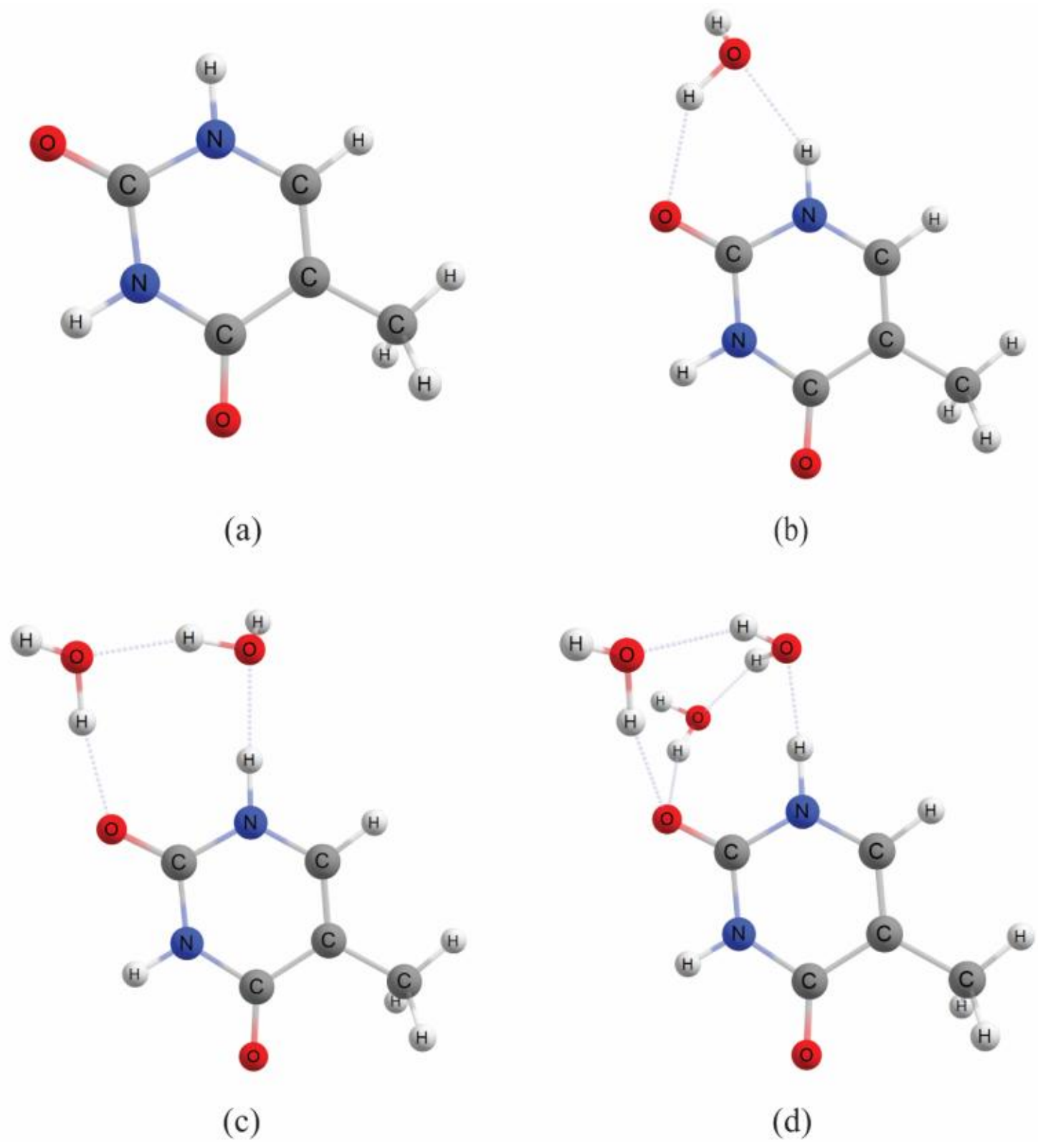

**Figure 1**. Optimized structures of thymine and microhydrated thymine [thymine($H_2O$)$_n$, n= 1, 2, 3] complexes at RI-MP2/def2-TZVP level of theory.

**Table 1.** Comparison of resonance positions (eV) and widths (eV) for gas-phase thymine obtained in this work with previous theoretical and experimental results.

| Resonance state | 1π* | | 2π* | | 3π* | |
|---|---|---|---|---|---|---|
| Method | $E_R$ | Γ | $E_R$ | Γ | $E_R$ | Γ |
| RVP-EA-EOM-DLPNO-CCSD (TIGHTPNO) | 0.67 | 0.017 | 2.42 | 0.06 | 5.38 | 0.12 |
| Projected CAP (box) EA-EOM-CCSD | 0.71 | 0.14 | 2.44 | 0.21 | 5.45 | 0.6 |
| GPA-EA-EOM-CCSD[47] | 0.68 | 0.02 | 2.32 | 0.073 | 5.02 | 0.58 |
| CAP/SAC-CI[75] | 0.67 | 0.11 | 2.28 | 0.15 | 5.14 | 0.41 |
| R-matrix/u-CC (2012)[59] | 0.53 | 0.08 | 2.41 | 0.10 | 5.26 | |
| R-matrix/SEP (2012)[59] | 0.60 | 0.11 | 2.73 | 0.11 | 5.52 | 0.57 |
| SMC/SEP (2007)[105] | 0.30 | | 1.9 | | 5.70 | |
| R-matrix/SE (2006)[104] | 2.40 | 0.20 | 5.50 | 0.60 | 7.90 | 1.00 |
| Expt.[103] | 0.29 | | 1.71 | | 4.05 | |

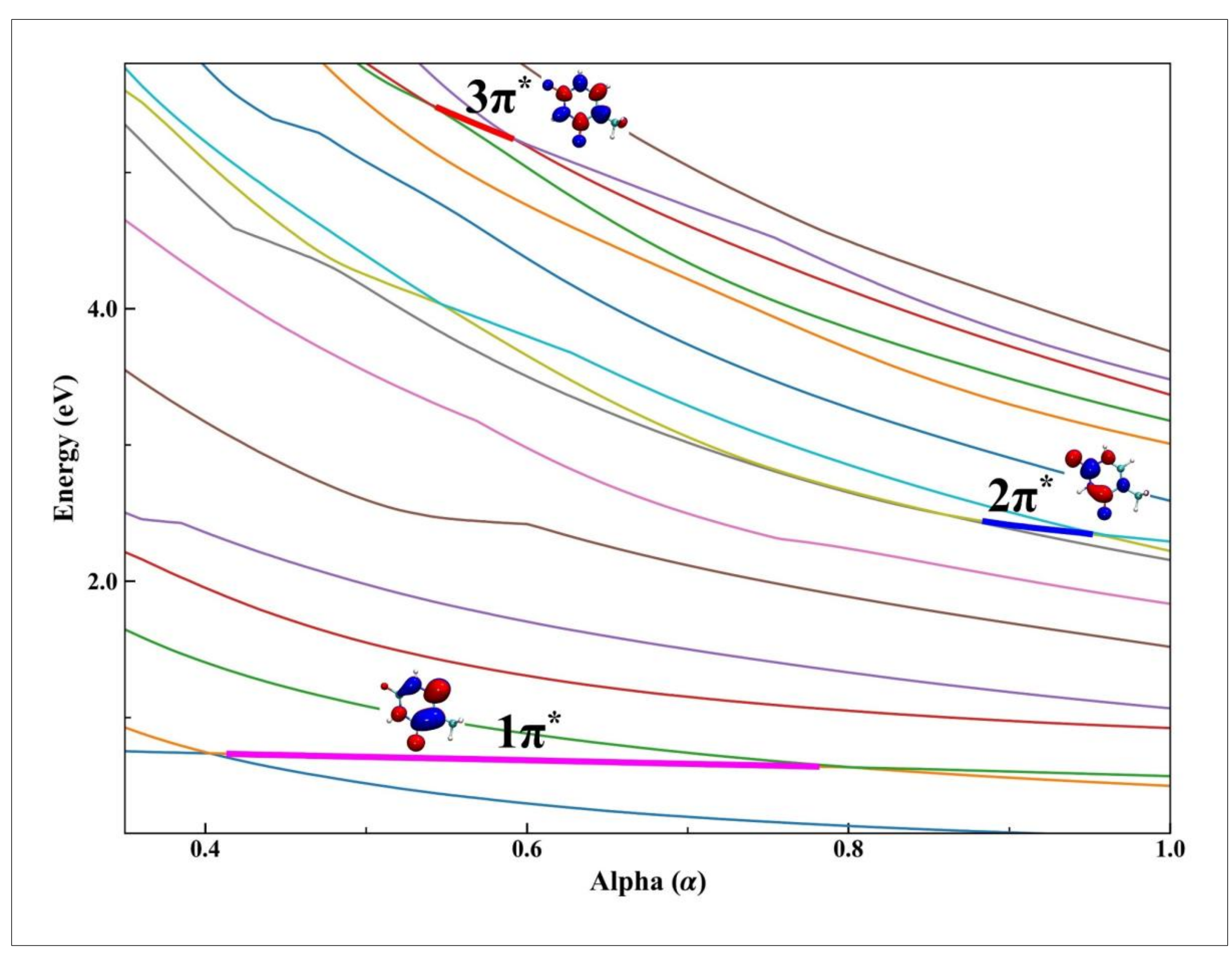

**Figure 2.** Stabilization plot of gas phase thymine at EA-EOM-DLPNO-CCSD/aug-cc-pVDZ* level of theory with TIGHTPNO setting.

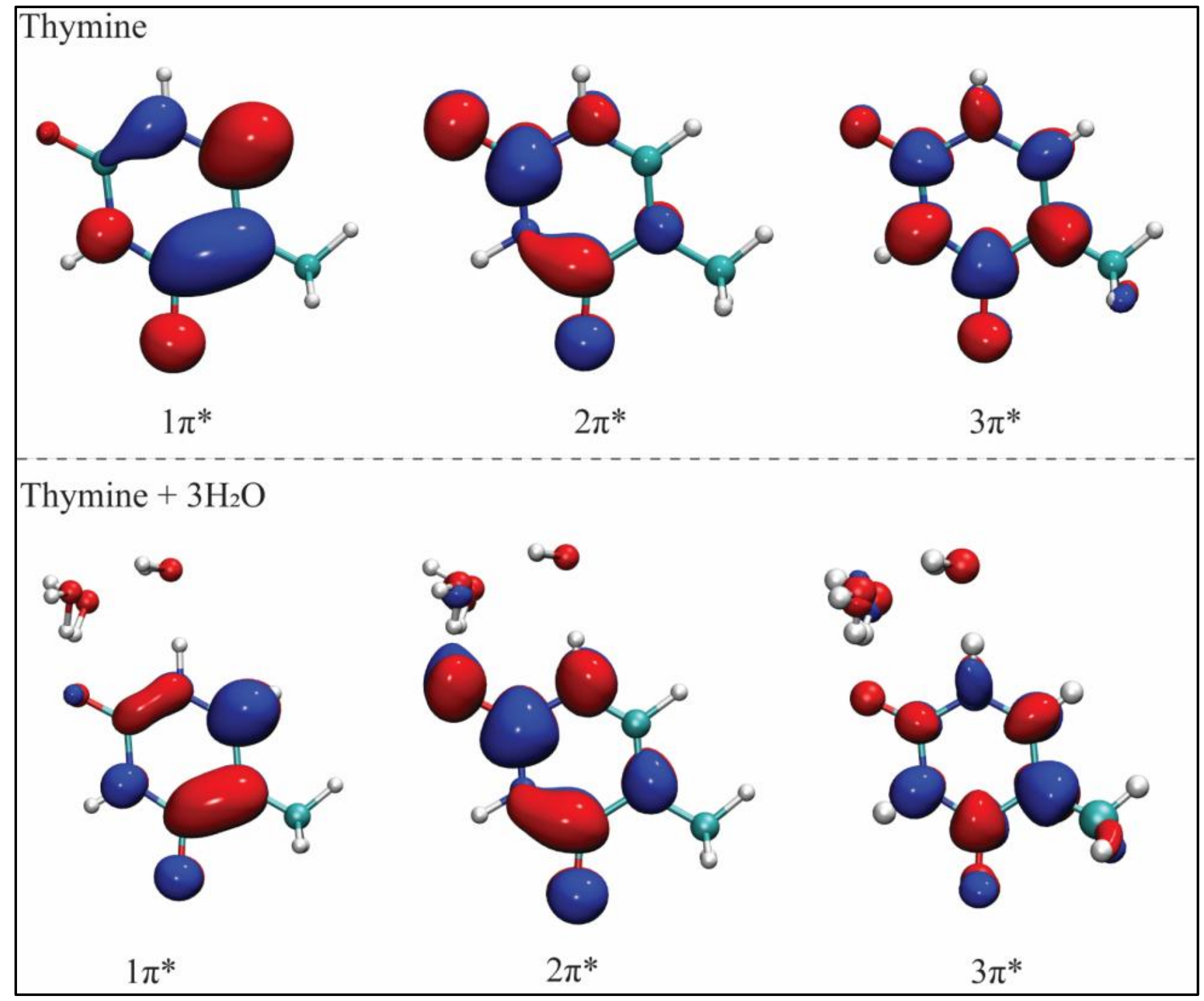

**Figure 3.** EA-EOM-DLPNO-CCSD natural orbitals corresponding to (a) 1π* (b) 2π* (c) 3π* shape-type resonance states of thymine and thymine$(H_2O)_3$ complex.

**Table 2.** Effect of microsolvation on the resonance positions and widths (in eV) of the 1π*, 2π*, and 3π* resonance states calculated using the RVP-EA-EOM-DLPNO-CCSD method with the TIGHTPNO setting and the aug-cc-pVDZ* basis set.

| Molecule | $E_R$ | Γ |
|---|---|---|
| thymine | 0.67 | 0.017 |
| | 2.42 | 0.06 |
| | 5.38 | 0.12 |
| thymine$(H_2O)_1$ | 0.66 | 0.016 |
| | 2.31 | 0.03 |
| | 5.46 | 0.157 |
| thymine$(H_2O)_2$ | 0.62 | 0.013 |
| | 2.12 | 0.024 |
| | 5.35 | 0.112 |
| thymine$(H_2O)_3$ | 0.53 | 0.006 |
| | 2.02 | 0.015 |
| | 5.09 | 0.084 |

**Table 3.** Effects of geometric distortion and finite basis set artifacts on the resonance positions and widths (in eV) of the 1π*, 2π*, and 3π* resonance states of thymine calculated using the RVP-EA-EOM-DLPNO-CCSD method with the aug-cc-pVDZ* basis set and the TIGHTPNO setting.

| Resonance state | 1π* | | 2π* | | 3π* | |
|---|---|---|---|---|---|---|
| System | $E_R$ | Γ | $E_R$ | Γ | $E_R$ | Γ |
| Isolated thymine in gas phase equilibrium geometry | 0.67 | 0.017 | 2.42 | 0.06 | 5.38 | 0.12 |
| Isolated thymine in thymine($H_2O$)$_1$ geometry | 0.69 | 0.02 | 2.45 | 0.061 | 5.69 | 0.23 |
| thymine($H_2O$)$_1$ with water as ghost | 0.67 | 0.018 | 2.43 | 0.058 | 5.57 | 0.19 |
| thymine($H_2O$)$_1$ | 0.66 | 0.016 | 2.31 | 0.03 | 5.46 | 0.157 |

**Table 4.** Effect of finite basis set artifacts on the resonance positions and widths (in eV) of the 1π*, 2π*, and 3π* resonance states of thymine($H_2O$)$_n$, (n = 2, 3) calculated using the RVP-EA-EOM-DLPNO-CCSD method with the aug-cc-pVDZ* basis set and the TIGHTPNO setting.

| Resonance state | 1π* | | 2π* | | 3π* | |
|---|---|---|---|---|---|---|
| System | $E_R$ | Γ | $E_R$ | Γ | $E_R$ | Γ |
| thymine($H_2O$)$_2$ with water as ghost | 0.67 | 0.016 | 2.33 | 0.033 | 5.42 | 0.12 |
| thymine($H_2O$)$_2$ | 0.62 | 0.013 | 2.12 | 0.024 | 5.35 | 0.11 |
| thymine($H_2O$)$_3$ with water as ghost | 0.61 | 0.016 | 2.27 | 0.032 | 5.46 | 0.15 |
| thymine($H_2O$)$_3$ | 0.53 | 0.006 | 2.02 | 0.015 | 5.09 | 0.08 |

**Table 5.** Influence of local hydrogen bonding geometry in thymine($H_2O$)$_1$ on the resonance positions and widths (in eV) of the 1π*, 2π*, and 3π* resonance states calculated using the RVP-EA-EOM-DLPNO-CCSD method with the aug-cc-pVDZ* basis set and the TIGHTPNO setting.

| Molecule | Conformer-0 | | Conformer-1 | | Conformer-2 | |
|---|---|---|---|---|---|---|
| thymine($H_2O$)$_1$ | $E_R$ | Γ | $E_R$ | Γ | $E_R$ | Γ |
| First resonance | 0.66 | 0.016 | 0.56 | 0.008 | 0.66 | 0.015 |
| Second resonance | 2.31 | 0.03 | 2.42 | 0.067 | 2.32 | 0.031 |
| Third resonance | 5.46 | 0.157 | 5.36 | 0.093 | 5.36 | 0.094 |

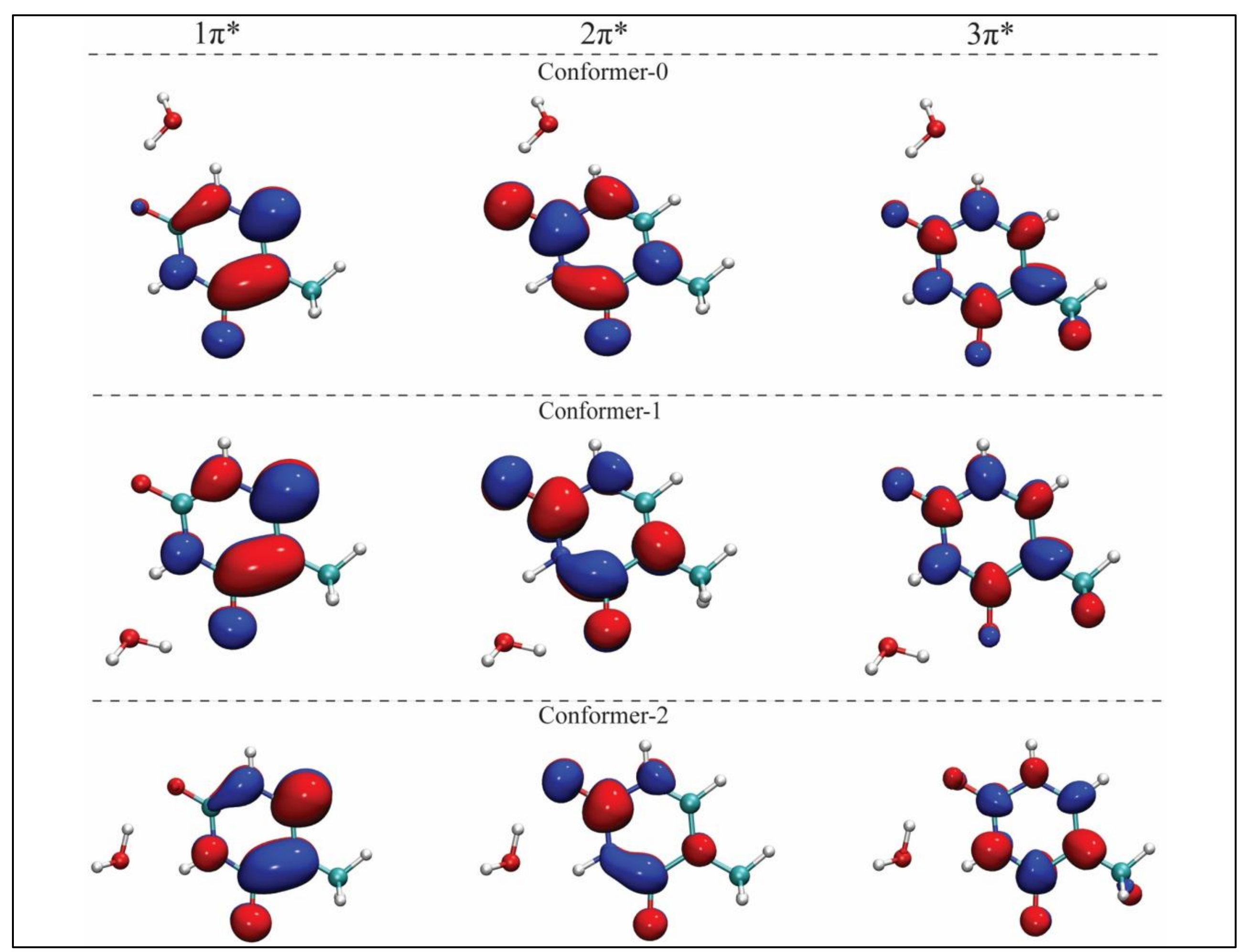

**Figure 4.** EA-EOM-DLPNO-CCSD natural orbitals corresponding to (a) 1π* (b) 2π* (c) 3π* shape-type resonance states of different thymine($H_2O$)$_1$ conformers.